\documentclass[letterpaper]{aa}
\usepackage{epsfig}
\usepackage{graphicx} 

\begin{document}

\title{Searching Gravitational Waves from Pulsars, Using Laser Beam
 Interferometers}

\author{Tania Regimbau \inst{1}
 \and Jose Antonio de Freitas Pacheco \inst{2}} 

\institute{LIGO laboratory, Massachusetts Institute of
 Technology, Cambridge, MA, 02139, USA
 \and Observatoire de la Cote d'Azur, BP 4229, 06304 Nice Cedex 4, France}

\offprints{Tania Regimbau,
 \email{tania@ligo.mit.edu}}

\date{Received \today / Accepted} 

\abstract{We use recent population synthesis results to investigate the 
distribution of pulsars in the frequency
space, having  a gravitational strain high enough to be
detected by the future generations of laser beam interferometers.
 We find that until detectors become able to recover the entire 
population, the frequency distribution of the 'detectable'
population will be very dependent on the
detector  noise curve. Assuming a mean equatorial deformation
$\varepsilon =10^{-6}$, the optimal frequency
is around 450 Hz for interferometers of the first generation (LIGO or
VIRGO) and shifts toward 85 Hz for advanced detectors.
An interesting result for future detection stategies is the significant 
narrowing of
the distribution when improving the sensitivity: with an advanced detector,
it is possible to have 90\% of detection probability 
while exploring less than 20\% of the 
parameter space (7.5\% in the case of $\varepsilon =10^{-5}$).
In addition, we show that in most cases the spindown of 'detectable' pulsars 
represents a period shift of less than a tens of nanoseconds after one year 
of observation, making them  easier to follow in the frequency space.

}

\titlerunning{Searching GW from pulsars...}
\maketitle

\section{Introduction}
The coming years will be marked by the beginning of operations for the
major gravitational wave interferometric antennas LIGO, VIRGO, GEO and
TAMA. Starting with a sensitivity of the order of
h $\approx 10^{-22}$, they are expected to evolve in a few years into
second generation experiments with a sensitivity improved
by one or two orders of magnitude depending on the frequency.   
Ground based detectors will scan the sky searching for gravitational waves 
with frequencies between  tens of hertz up to tens
of kilohertz. Potential sources fall roughly into three classes:
bursts, stochastic background, continuous, involving different
search techniques as match filtering for coalescing binary systems and
cross-correlation between detectors for the stochastic background. The 
detection of pulsars (and other nearly monochromatic continuous sources) 
can be achieved by integrating the signal during times of about $10^{7}$ s 
and searching for statistically significant peaks at fixed frequencies in 
the power spectrum.   
This method becomes quite complicated as soon as one consider that:
1) pulsars are not really monochromatic emitters. The Doppler effect due to
the Earth motion and the intrinsic spindown of the star become 
significant over long period of times spreading power across many 
frequency bins and must be corrected before performing the fourier transform.  
2) most of pulsars are not observed in the electromagnetic domain due
to strong selection effects, making necessary an all-sky and all-frequency
search. 
The computational cost is therefore the main concern of searching for
continuous sources: if the whole parameter space is scanned, the 
computing power required
to process the data will be out of any reasonable value: of the order of
$3 \times 10^{20}$ Tera-flops for integration times of one year,
frequency bandwiths of 2500Hz and decay times of 1000 year (Frasca et
al. 2000). Alternative search algorithms involving hierarchical methods which
follow up candidate detections from a first pass search have been
investigated during the past few years (Brady et al. 2000, 
Schutz \& Papa 1999, among others). 
These techniques reducing very significantly the computing time, by more 
than ten orders of
magnitude for a small loss of sensitivity (by less than an
order of magnitude) (Frasca et al. 2000) represent the best hope of
detection for actual computational ressources and detector sensitivities.
An attempt to optimize these algorithms could be the determination of
the most likely range on the parameter space. Without restricting the 
search area definitively,
the probability to have a first detection early could be optimized.
Brady et al. (1998) have considered a survey in a limited sky area, 
searching for 
pulsars in the galactic core.
In a previous work (Regimbau \& de Freitas Pacheco 2000, hereafter paper I), 
population synthesis of galactic normal
pulsars was performed in
order to recover the statistical properties of the real population. 
This approach permits the modeling of the galactic pulsar
contribution to the gravitational emission and to deduce its
detectability by laser beam interferometers.
The gravitational strain being proportional to the star equatorial
deformation, the number of detection depends on the mean value adopted
for the equatorial ellipticity. 
This parameter is quite uncertain since it depends on the cooling history 
and on different
strains underwent by the crust. Past studies (Thorne 1980) have shown
that equatorial distortions cannot be larger than $10^{-5}$, expected
to be an upper limit related to the maximum stress that the solid
crust could support, but it is worth mentioning that it could be
several orders of magnitude smaller. 
Our simulations have showed that $10^{-6}$ is the critical value to have 
at least one
detection with interferometers of the first generation
($10^{-7}$ for an advanced interferometer)
In this paper, by using our previous population synthetis we investigate
the detection probability of a pulsar in the frequency space.
In order to illustrate our
findings, we report here results for mean ellipticities equal 
to $10^{-5}$ and
$10^{-6}$, as well as how the results are affected by the
detector sensitivity.

\section{The Population Synthesis}

\subsection{The Model}

Population synthesis aimed to estimate the gravitational contribution
of galactic normal pulsars are described in details in paper I.
The initial parameter distributions being fully taken into
account, our statistical approach gives more realistic estimates of
the gravitational strain and the number of 'potentially detectable'
pulsars with frequencies falling into the detector sensitivity band, than
other analytical methods.
Among the main results of our simulations those relevants for the
present purpose are briefly reminded here: 
a) the average initial period of pulsars is 290 ms with a dispersion
of 100 ms. Pulsars are not born in majority as fast rotators in agreement with
recent studies (Bhattacharya et al. 1992, Hartman et al. 1997, Narayan
1987) and with observations, since the mean
period of pulsars younger than 20, 50 and 100 kyr is
respectively 123, 162 and 173 ms (Lyne \& Granham-Smith 1998). However, the 
present simulations with a non-fixed initial
period permit to account for the fastest objects like the Crab 
pulsar (33.4 ms) and for the diversity of  young pulsar
periods: PSR B1509-58 (as young as the Crab pulsar) has a period of
150 ms, PSR B1610-50 (six times older) a period of 231 ms, while the
VELA pulsar and PSR B1951+32, respectively 10 and 100 times older than
the Crab pulsar rotate with periods of 89 ms and 59 ms. 
b) around 150000 pulsars are expected to be present in the Galaxy. 
c) among them, pulsars in the detector frequency range of sensitivity (called
later "potentially detectables") would be around 1000 if the lower
bound is 10 Hz or around 5200 if it is 5 Hz. The latter  is
expected to be attained in a second phase of the VIRGO experiment,
thanks to a sophisticated pendulum system called
superatenuator. Notice that the upper bound of the detector (around
5-10 kHz) doesn't interpose on the detection of pulsars, since the 
Keplerian stability limit of neutron stars corresponds to gravitational wave
frequencies of 2-4 kHz, depending on the equation of state.

\subsection{The Detectable Population}

When any precessional effect is neglected  the two
polarization components of gravitational waves emitted by a rotating
neutron star are (Bonazzola \& Gourgoulhon 1996)
\begin{equation}
h_{+}(t)=2A(1+\cos ^{2}i)\cos (2\Omega t)
\end{equation}
and 
\begin{equation}
h_{\times }(t)=4A\cos i\sin (2\Omega t)
\end{equation}
where $i$ is the angle between the spin axis and the wave propagation
vector, assumed to coincide with the line of sight, and the wave
amplitude is defined by 
\begin{equation}
A=\frac{G}{rc^{4}}\varepsilon I_{zz}\Omega ^{2}
\end{equation}
In this expression, G is the gravitation constant, c is the velocity of 
light, r is the distance to the source, $\Omega $ is the angular rotation
velocity of the pulsar and the ellipticity $\varepsilon $ is defined as 
\begin{equation}
\varepsilon =\frac{I_{xx}-I_{yy}}{I_{zz}}
\end{equation}
The gravitational strain amplitude induced in the detector depends
also on the direction and polarization of the wave with relative  to the
detector arms and it is given by:
\begin{equation}
h(t)=h_{+}(t)F_{+}(\theta ,\phi ,\psi )+h_{\times }(t)F_{\times }(\theta,
\phi ,\psi )
\end{equation}
where $F_{+,\times}$ are the beam factors of the
interferometer (Jaranowski et al. 1998), which are functions of the zenith
distance $\theta $, the azimuth $\phi $ as well as of the wave
polarization plane orientation $\psi $.
 
Among the ''potentially detectable'' population mentioned above
(pulsars with periods less than 0.4 s for VIRGO and less than 0.2 s
for LIGO), 'detectable' objects are pulsars with gravitational
strain amplitudes above the antenna sensitivity curve.
The detectability criterium can be expressed as follows:
\begin{equation}
h(f_{g})\geq 2\frac{h_{n}(f_{g})}{\sqrt{T}}
\end{equation}
where $h$ is the gravitational strain, $T$ the integration time,
$h_{n}$ the detector
sensitivity (in Hz$^{-1/2})$. In our analysis, the planned LIGO II was
used as an exemple of second generation detectors. The factor two 
corresponds to the
adopted signal-to-noise ratio. 
As one should expect, the less is $\varepsilon $, the harder is the 
detectability
condition. The number of detections as a function of the ellipticity
and detector sensitivity is reported in Table 1.
An average ellipticity of $10^{-6}$ represents the detectability
threshold for detectors of the first
generation as VIRGO or LIGO I: around $12-15$ detections are 
expected if $\varepsilon
=10^{-5}$, $2-3$ detections if $\varepsilon =10^{-6}$ and no detection
if $\varepsilon <10^{-6}$. 
For an advanced interferometer this threshold corresponds to a mean
ellipticity of $10^{-7}$. 
This result has quite interesting consequences: even
in the case of no detection, an upper limit can be set to the ellipticity 
and improvements in the
sensibility will gradually push down such a limit.

\begin{table}
\caption{Number of detections as a function of the
  ellipticity for interferometers of the first generation (LIGO and
  VIRGO) and for an advanced interferometer}
\begin{flushleft}
\begin{tabular}{lccc}
\noalign{\smallskip}
\hline
\noalign{\smallskip}
$\varepsilon $ & $ 10^{-5} $ & $ 10^{-6} $ & $ 10^{-7} $ \\
\noalign{\smallskip}
\hline
\noalign{\smallskip}
VIRGO & 15 & 3 & 0\\
LIGO I & 12 & 2 & 0\\
Advanced & 90 & 12 & 2\\
\noalign{\smallskip}
\hline
\end{tabular}
\end{flushleft}
\end{table}

As shown in Table 1, detectable pulsars are expected
to be few at the beginning, since only the nearest and the fastest
objects have strain amplitudes high enough to be above the noise level. 
The detector sensitivity acts as a filter
affecting a priori both spatial and frequency statistics.
Actually our simulations have shown that the distance distribution is
quite similar for the detectable population and the total population, 
with a peak around the galactic
center. As a result the sky direction distributions do not depend
significantly on the antenna sensitivity or on the ellipticity. On the
opposite, the frequency distribution is very dependent on both of them
and requires a more detailed analysis.

\subsection{The Gravitationnal Frequency}

The gravitational frequency distribution of detectable pulsars as a
 function of the antenna sensitivity,
 for two different values of the
ellipticity, the theoretical upper value $\varepsilon =10^{-5}$ 
and the detectability limit for interferometers of the first
generation, $\varepsilon =10^{-6}$ are shown respectively in
Fig. 1 and Fig. 2.

\begin{figure}
 \resizebox{\hsize}{!}{\includegraphics[angle=-90]{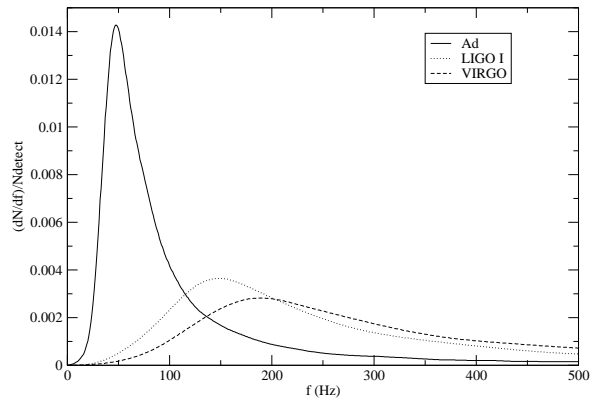}}\hfill
\caption{frequency distribution of detectable pulsars 
if $\varepsilon =10^{-5}$}
\end{figure}

\begin{figure}
\resizebox{\hsize}{!}{\includegraphics[angle=-90]{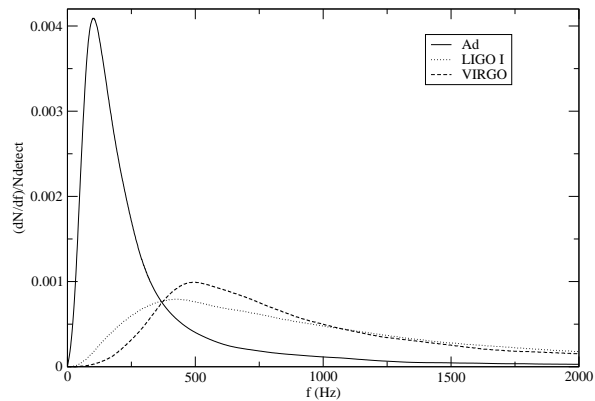}}\hfill
\caption{frequency distribution of detectable pulsars 
if $\varepsilon =10^{-6}$}
\end{figure}

The sensitivity improvement enables the detection of slower objects, 
so the maximum of the detectable frequency distribution shifts toward lower 
values. For $\varepsilon =10^{-5}$, the maximum is respectively around 
180 Hz and 135
Hz for VIRGO and LIGO I and decreases to 85 Hz for an advanced interferometer. 
For $\varepsilon =10^{-6}$ it is 480 Hz for VIRGO, 435 Hz for LIGO I
and 85 Hz for for an advanced interferometer. The difference between
the optimal frequencies of the two detectors of the first generation
arises from the fact that LIGO I is more sensitive than VIRGO between 50 Hz and
330 Hz and less sensitive elsewhere.
What is interesting to notice for future detection
stategies, is the significant reduction of the distribution width 
between the first and second generation (see Table 2). It 
decreases from 200 Hz to 50 Hz if $\varepsilon =10^{-5}$ and 
from 700-1000 Hz to 200 Hz if $\varepsilon
=10^{-6}$. Consequently, we could keep 90\% of the maximum detection 
probability while exploring less than 7.5\% of the frequency space
with an advanced detector versus 25\% with interferometers of the first
generation. Another important result concerns the lower bound of the 
detector frequency band. At the begining, this value doesn't play much 
part but the improvement
of the sensitivity would step by step allow to recover the all
''potentially detectable'' population concentrated at the lowest frequencies. 
\begin{table}
\caption{parameters of the detectable frequency distribution for
$\varepsilon =10^{-5}$ (left value) and $\varepsilon =10^{-6}$(right
value) for interferometers of the first generation (LIGO and
  VIRGO) and for an advanced interferometer: the maximum, the width and 
the frequency band around the
maximum where we find 90\% of detection}
\begin{flushleft}
\begin{tabular}{lccc}
\noalign{\smallskip}
\hline
\noalign{\smallskip}
&max (Hz) & width (Hz) & 90\% band (Hz) \\
\noalign{\smallskip}
\hline
\noalign{\smallskip}
VIRGO & 180 - 480 &  200 - 700 & [0-1200]-[0-2500]\\
LIGO I & 135 - 435 & 200 - 1000 &[0-1000]-[0-2600]\\
Advanced & 45 - 85 &  50 - 200 & [0-300]-[0-2500] \\
\noalign{\smallskip}
\hline
\end{tabular}
\end{flushleft}
\end{table}

As already mentionned one of the major complication when searching for
pulsars over long period of time arises from the intrinsic spindown
which spreads power across many frequency bins. 
A model of the intrinsic frequency evolution is thus needed to
remove these effects before performing the Fourier transform.
Estimates of the spindown parameters can be derived
from observations of radio pulsars (Brady et al., 1998 and 2000 for 
instance) but strong selection effects  make them non representative of 
the detectable population.
In the following, we investigate the statistical properties of the 
spindown rate of detectable pulsars. The period derivative 
distribution of detectable pulsars as a
function of the antenna sensitivity and for an 
ellipticity $\varepsilon =10^{-6}$ is shown in
Fig.3 and compared with the expected distribution of
the 'potentially detectable' population. 
As there is no significant differences between the distributions of
the first generation interferometers, only one plot is reported for both LIGO I
and VIRGO.
\begin{figure}
\resizebox{\hsize}{!}{\includegraphics[angle=-90]{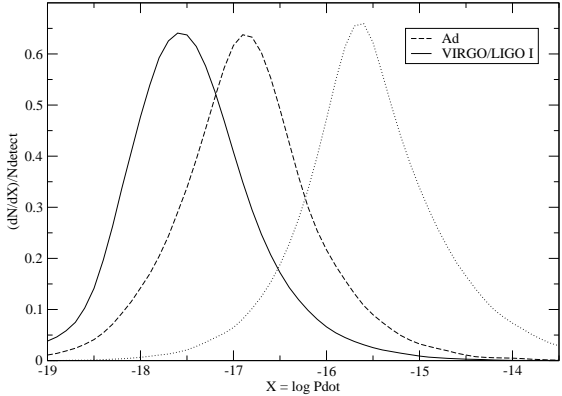}}\hfill
\caption{period derivative distribution of the detectable pulsar population if
  $\varepsilon =10^{-6}$ for the sensitivity of interferometers of the
  first generation (plain line) as well as for an advanced
  interferometer (dashed line). The dot line represents the expected
  distribution of the 'potentially' detectable population}
\end{figure}

The period derivative probability function can be represented quite
well by a gaussian distribution with standard deviation $\sigma=0.6$. The
maximum (or the mean value) is around $\log(\dot{P})=-17.5$ for interferometers
of the first generation and increases to $\log(\dot{P})=-16.8$ for an advanced
interferometer. As the sensitivity improves, the distribution shifts
toward the expected distribution of the 'potentially detectable'
population which maximum is  $\log(\dot{P})=-15.6$.
Changing the mean ellipticity
from $\varepsilon =10^{-6}$ to $\varepsilon =10^{-5}$ acts the same
way (see Table 3)
The general belief is that detectable pulsars will be the youngest at 
the beginning of the operations  but it is shown here that it is not 
necessarily the case. Actually, detectable
pulsars are in majority the 'most stable'. That is very understandable
considering that the less is its spindown, the longer a pulsar remain in 
the high frequency range, which increase
its probability to be detected.  

It is relevant to define the critical spindown rate $\dot{P}_{90\%}$
below which 90\% of detectable pulsars can be found. This parameter
may be used to determine the number of bins that
should be investigated in the frequency  space to correct spindown
effects.  
Table 3 gives $\dot{P}_{90\%}$ as well as a more conservative
$\dot{P}_{99\%}$, for interferometers of the first and second
generations and for ellipticities $\varepsilon =10^{-5}$ and $10^{-6}$.
Columns $\Delta P_{90\%}$ and $\Delta P_{99\%}$ give an estimation 
of the maximal shift one should expect (at 90\% and 99\% of
confidence) after one year of observation (equation 7 with $T=1$ yr)  
\begin{equation}
\Delta P_{90\%}=\dot{P}_{90\%}\times T 
\end{equation}

\begin{table}
\caption{parameters of the period derivative distribution for
$\varepsilon =10^{-6}$ (left value) and $\varepsilon =10^{-5}$(in
parenthesis) for interferometers of the first generation and for an
advanced interferometer: the maximum, the 0.9 and 0.99 quantiles
(lines 3 \& 5) and 
the and the corresponding period shifts in nanosecond after one year of 
observation (lines 4 \& 6)}
\begin{flushleft}
\begin{tabular}{lccc}
\noalign{\smallskip}
\hline
\noalign{\smallskip}
&LIGO/VIRGO & Advanced \\
\noalign{\smallskip}
\hline
\noalign{\smallskip}
max &-17.5 (-17.2)& -16.8 (-16.4) \\
$log \dot{P}_{90\%}$&-16.7 (-16.4) &-16.0 (-15.6)\\
$\Delta P_{90\%}$ (ns) &0.63 (1.26) &3.15 (7.93)\\
$log \dot{P}_{99\%}$&-16.1 (-15.8) &-15.4 (-15)\\
$\Delta P_{99\%}$(ns)&2.51 (5.00) &12.65 (31.56)\\
\noalign{\smallskip}
\hline
\end{tabular}
\end{flushleft}
\end{table}

\subsection{Conclusions}

In this paper, the distribution in the frequency 
space, of pulsars with a gravitational strain high enough to be
detected by the future generations of laser beam interferometers was 
investigated.
Until detectors become able to recover the entire population, the
'detectable' frequency distribution  will be very dependent on the
detector sensitivity. If $\varepsilon =10^{-6}$, the optimal frequency is 
around 450 Hz for interferometers of the first generation and shifts
toward 85 Hz for an advanced detector. An encouraging result for 
future detection stategies is the significant narrowing of
the optimal band of frequency which follows the improvement of the
sensitivity: with an advanced detector we could keep 90\% of the
maximum detection probability while exploring less than 20\% of the
frequency space (7.5\% $\varepsilon =10^{-5}$).
In addition, we showed that in most cases 'detectable' pulsars have very 
stable rotations, making them easier to track in the frequency space. The 
critical spindown rate below which 90\% of detectable pulsars can be found 
is of the order of a few nanosecond per year for interferometers of the 
first generation (and a tens of nanosecond per year for an 
advanced interferometer).
These informations, integrated into detection algorithms may be usefull to 
choose the optimal way to start blind searches, increasing our chances to 
detect these objects.


\begin{thebibliography}{}
\bibitem[1192]{Bhattacharya}
Bhattacharya, D., Wijers R., Hartman, J.W., \& Verbunt, F. 1992, A\&A 254, 198
\bibitem[1996]{Bonazzola}
Bonazzola, S., \& Gourgoulhon, E. 1996, A\&A, 312, 675
\bibitem[1998]{Brady1}
Brady, P.R., Creighton, T., Cutler, C., \& Schutz, B.F. 1998,
Phys.Rev.D, 57,4
\bibitem[2000]{Brady2}
Brady, P.R., \& Creighton, T. 2000, Phys.Rev.D, 61, 8 
\bibitem[2000]{Frasca}
Frasca, S. 2000, Int. J. Mod. Phys. D 9, 369
\bibitem[1997]{Hartman}
Hartman, J.W., Bhattacharya, D., Wijers, R., \& Verbunt, F. 1997, A\&A
322, 477
\bibitem[1998]{Jaranowski}
Jaranowski, P., Kr\'olak, A.,\& Schutz, B.F. 1998, gr-qc/9804014
\bibitem[1998]{Lyne}
Lyne, A.G., \& Graham-Smith, F. 1998. In Pulsar Astronomy, Cambridge
University Press, Cambridge
\bibitem[1987]{Narayan}
Narayan R. 1987, ApJ, 319, 162
\bibitem[2000]{Regimbau}
Regimbau, T., \& de Freitas Pacheco,J.A. 2000, A\&A, 359, 242, Paper I
\bibitem[2000]{Schutz}
 Schutz, B.F., \& Papa, M.A. 1999, gr-qc/9905018 
\bibitem[1980]{Thorne}
Thorne, K.S. 1980, Rev.Mod.Phys. 52, 299
\bibitem[1979]{Zimmermn}
Zimmerman, M., \& Szedenits, E. 1979, Phys.Rev.D 20, 351 



\end{thebibliography}
\end{document}